\documentclass[12pt]{article}

\usepackage{amsmath,amsthm,amssymb}
\usepackage{graphicx,psfrag,epsf}
\usepackage[caption=false]{subfig}
\usepackage[dvipsnames]{xcolor}
\usepackage{enumerate}
\usepackage{adjustbox,multirow,booktabs}
\usepackage{url} 
\usepackage{natbib}

\usepackage{geometry}
\newtheorem{theorem}{Theorem}
 
\DeclareGraphicsExtensions{.png,.pdf}

\providecommand{\keywords}[1]{\textbf{Keywords:} #1}

\title{On the bias of H-scores for comparing biclusters, \\ and how to correct it}

\author{
	Jacopo Di Iorio\thanks{jacopo.diiorio@polimi.it} \\
	\small{MOX - Dept.~of Mathematics, Politecnico di Milano. Milano, Italy}
	\and
	Francesca Chiaromonte\thanks{fxc11@psu.edu} \\ 
	\small{Dept.~of Statistics, Penn State University. University Park, PA USA} \\
	\small{Inst.~of Economics and EMbeDS, Sant’Anna School of Advanced Studies. Pisa, Italy}
	\and
	Marzia A. Cremona\thanks{mac78@psu.edu} \\ 
	\small{Dept.~of Statistics, Penn State University. University Park, PA USA}
}

\begin{document}
\maketitle

\abstract{ 
	In the last two decades several biclustering methods have been developed as new unsupervised learning techniques to simultaneously cluster rows and columns of a data matrix. 
	These algorithms play a central role in contemporary machine learning and in many applications, e.g. to computational biology and bioinformatics. 
	The H-score is the evaluation score underlying the seminal biclustering algorithm by Cheng and Church, as well as many other subsequent biclustering methods. 
	In this paper, we characterize a potentially troublesome bias in this score, that can distort biclustering results. 
	We prove, both analytically and by simulation, that the average H-score increases with the number of rows/columns in a bicluster.
	This makes the H-score, and hence all algorithms based on it, biased towards small clusters. 
	Based on our analytical proof, we are able to provide a straightforward way to correct this bias, allowing users to accurately compare biclusters.
}


\section{Introduction}
The \emph{H-score} (or Mean Squared Residue score, MSR) underlies Cheng and Church’s biclustering algorithm (\citeyear{cheng2000}), one of the best-known and most widely employed algorithms in bioinformatics and computational biology, and many subsequent algorithms (e.g., FLOC, \citealt{yang2005improved}, and \citealt{huang2011parallelized}). 
Cheng and Church’s algorithm has \textasciitilde2400 citations to date, 597 since 2015, and 179 in 2018-19 alone. 
It was the first to be applied to gene microarray data, and it is one of the main tools available in biclustering packages (e.g., the “biclust” R library) as well as in gene expression data analysis packages (e.g., IRIS-EDA, \citealt{Monier2019IRIS}). 
In addition, it is widely used as a benchmark: almost all published biclustering algorithms include a comparison with it. 
The role of the H-score in a biclustering algorithm  
is to allow validation and comparisons of biclusters, which may have different numbers of rows and columns. 
Our findings document a bias that can distort biclustering results. We prove, both analytically and by simulation, that the average H-score 
increases with the number of rows/columns in a bicluster -- 
even in the “ideal” (and simplest) case of a single bicluster generated by an additive model plus a white noise. 
This biases the H-score, and hence all H-score based algorithms, towards small biclusters. 
Importantly, our analytical proof provides a straightforward way to correct this bias. 

\section{H-scores as a measure of bicluster coherence}
\citet{cheng2000} were the first to introduce biclustering as a way to identify (possibly overlapping) subsets of genes and/or conditions showing high similarity in a gene expression data matrix.
The H-score they proposed to measure (dis)similarity is defined as
"the variance of the set of all elements in the bicluster, plus the mean row variance and the mean column variance". 
Unlike measures employed by traditional clustering algorithms, the H-score is not a function of pairs of genes or conditions, but rather a quantitation of the coherence of all genes and conditions within a bicluster. 
Let $A=(a_{ij})$ be a data matrix. 
The H-score of the submatrix identified by the pair of index subsets $(I,J)$ 
is defined as
\begin{equation*}
    H(I,J) = \frac{1}{\mid I \mid \mid J \mid} \sum_{i \in I, j \in J} \left( a_{ij} - a_{iJ} - a_{Ij} + a_{IJ} \right)^{2}
\end{equation*}
where $a_{iJ}$, $a_{Ij}$ and $a_{IJ}$ are the means of row $i$, column $j$, and 
of the whole submatrix $(I,J)$, respectively.

An optimal bicluster is a submatrix 
$(I,J)$ 
with the lowest possible H-score $H(I,J)=0$ \citep{madeira2004biclustering}. 
This corresponds to a bicluster 
perfectly 
defined by the additive model $a_{ij} = \mu + \alpha_{i} + \beta_{j}$, where $\mu$ is the 
mean of the bicluster, and $\alpha_{i}$ and $\beta_{j}$ are additive adjustments for rows and columns, respectively. 
As an example, in the case of gene expression data, 
such a bicluster is a group of genes with expression levels that tend to fluctuate in unison across a group of conditions. 
In general, an additive error term $\epsilon_{ij}$ is also present, leading to the $H(I,J)>0$ 
for the model $a_{ij} = \mu + \alpha_{i} + \beta_{j} + \epsilon_{ij}$. 

The algorithm proposed by \citet{cheng2000} 
starts from the entire matrix $A$, iteratively deletes rows and/or columns which contribute to the H-score the most, and stops when the the current submatrix has $H(I,J)<\delta$ -- a given threshold. 
This identifies a so-called $\delta$-bicluster. 
To find additional $\delta$-biclusters, the procedure is repeated after replacing the entries of the prior $\delta$-bicluster(s)
with random numbers. 
Following \citet{cheng2000}, many other biclustering algorithms based on the H-score have appeared in the literature. 
For example, \citet{yang2005improved} proposed FLOC, a probabilistic algorithm that simultaneously identifies a set of $k$ (possibly overlapping) biclusters with low H-score. The procedure iteratively reduces the H-scores of $k$ randomly initialized biclusters, until the overall biclustering quality stops improving. 
\citet{angiulli2008random} proposed an algorithm based on a greedy technique combined with a local search strategy to escape poor local minima. This also employs the H-score, together with the row (gene) variance and the size of the bicluster. 
The Reactive GRASP Biclustering  \citep[RGRASP-B,][]{dharan2009biclustering} also uses the H-score 
to evaluate bicluster quality, and the algorithm in \citet{bryan2006application} uses a modified version of it. 
The algorithms cited here are only a small subset of those that rely on the H-score 
to validate and evaluate results 
\citep[see e.g.,][for an extensive review]{pontes2015biclustering}.

\begin{figure}[!tb]
    \centering
    \subfloat[\label{subfig:Hscore}]{\includegraphics[width=0.45\textwidth]{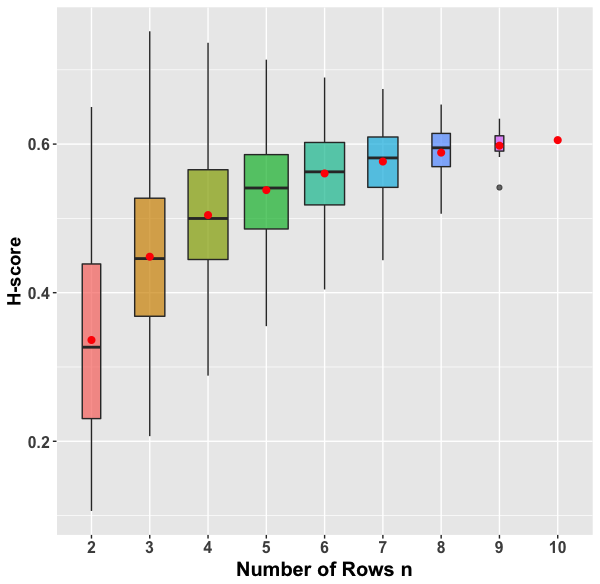}} \\
    \subfloat[\label{subfig:ratio}]{\includegraphics[width=0.45\textwidth]{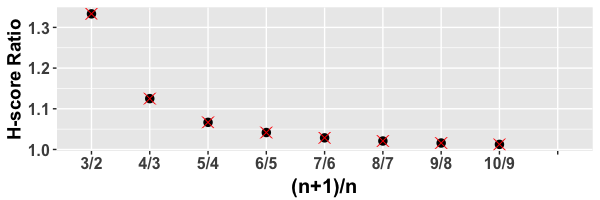}}
    \caption{
        \protect\subref{subfig:Hscore} H-score as function of the number of rows (columns), in a single bicluster of 10 columns (rows) generated by an additive model plus a white noise 
        $a_{ij} = \mu + \epsilon_{ij}$. 
        Red dots indicate average H-scores $\overline{H}_{n}$. 
        \protect\subref{subfig:ratio} Ratio $r_{n,n+1}$ between average H-scores, with $n=2,\dots,9$ according to simulation (black dot) and Theorem 1 (red cross). 
    }
    \label{fig:example}
\end{figure}

\section{Bias of H-scores for bicluster comparison}
Consider a bicluster $(I,J)$ generated by 
the additive model 
\mbox{$a_{ij} = \mu + \alpha_{i} + \beta_{j} + \varepsilon_{ij}$}. 
As mentioned above, the error  
$\varepsilon_{ij}$, 
renders $H(I,J) > 0$. 
However, 
in addition to the amount of noise, the H-score depends also on {\em size}, i.e. on the number of rows and columns in the bicluster. 
%
Figure \ref{fig:example}\protect\subref{subfig:Hscore} shows that the average H-score $\overline{H}_{n}$ ($\overline{H}_{p}$) of all 
possible 
submatrices of 
the bicluster 
$(I,J)$ having fixed number of columns (rows) and $n$ rows ($p$ columns), increases with $n$ ($p$). 
In particular, the relationship between $\overline{H}_{n}$ and $\overline{H}_{n+1}$ is expressed by the following Theorem, which proof can be found in the Appendix. An analogous result hold for $\overline{H}_{p}$ and $\overline{H}_{p+1}$. 

\begin{theorem}
    Let $(I,J)$ be a bicluster generated by the additive model $a_{ij} = \mu + \alpha_{i} + \beta_{j} + \varepsilon_{ij}$. 
    Let $\overline{H}_{n}$ be the average H-score of all the possible sub-matrices of $(I,J)$ having $n=2,3,\dots $ rows and a fixed number of columns $p$. 
    Then
    \begin{equation}
        \overline{H}_{n+1} = \overline{H}_{n} \frac{n^2}{n^2-1}.
    \label{eq:ratio}
    \end{equation}
    Analogously, we have 
    \begin{equation}
    	\overline{H}_{p+1} = \overline{H}_{p} \frac{p^2}{p^2-1},
    \label{eq:ratio_bis}
    \end{equation}
    with $\overline{H}_{p}$ the average H-score of all the possible sub-matrices of $(I,J)$ having a fixed number of rows $n$ and $p=2,3,\dots $ columns.
\end{theorem}


Focusing on rows (an identical reasoning holds for columns), we thus have that the ratio $r_{n,n+1} = \frac{\overline{H}_{n+1}}{\overline{H}_{n}}$ is fully determined by the number of rows $n$. 
From 
Theorem 1
it also follows that
\begin{equation}
    \overline{H}_{n+m} = \overline{H}_{n}\prod_{i=n}^{n+m-1}\frac{i^2}{i^2-1}.
\label{eq:evo}
\end{equation}
and therefore that knowing $\overline{H}_{n}$ is sufficient to compute $\overline{H}_{n+m}$ for every \mbox{$m>0$}. 
Since $r_{n,n+m} = \frac{\overline{H}_{n+m}}{\overline{H}_{n}}$, for $m\to\infty$ we obtain 
\begin{equation}
    r_{n,n+m}=\frac{\overline{H}_{n+m}}{\overline{H}_{n}}
    \longrightarrow\prod_{i=n}^{\infty}\frac{i^2}{i^2-1}.
\label{eq:infprod}
\end{equation}

\begin{figure}[!hb]
	\centering
	\subfloat[\label{subfig:HscoreN1}]{\includegraphics[width=0.46\textwidth]{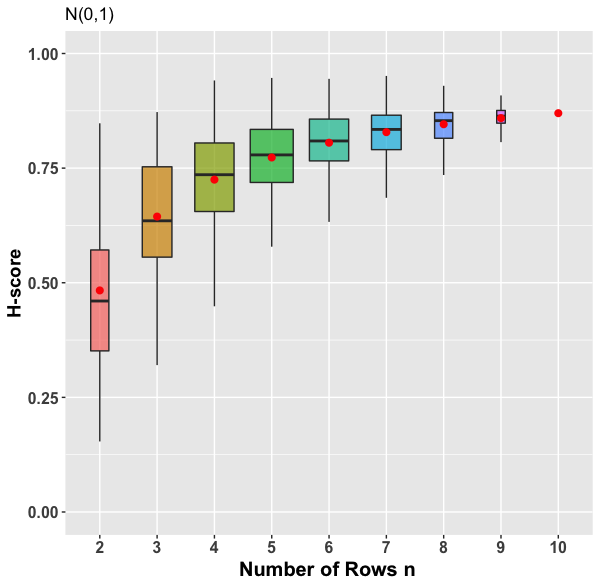}} \hspace{1cm}
	\subfloat[\label{subfig:HscoreU1}]{\includegraphics[width=0.46\textwidth]{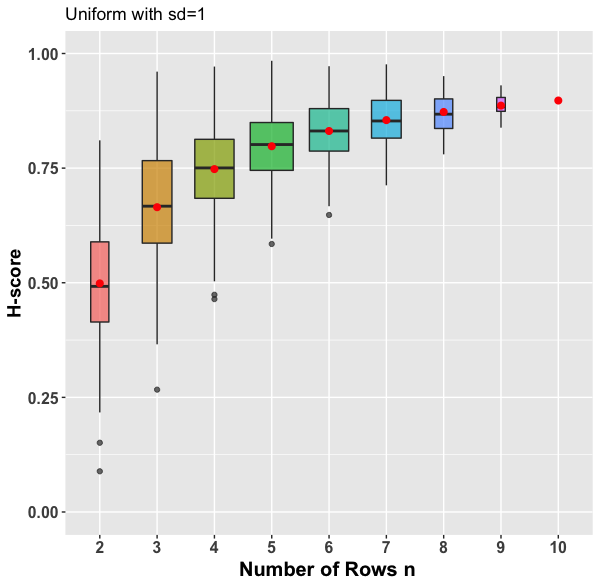}} \\
	\subfloat[\label{subfig:ratioN1}]{\includegraphics[width=0.46\textwidth]{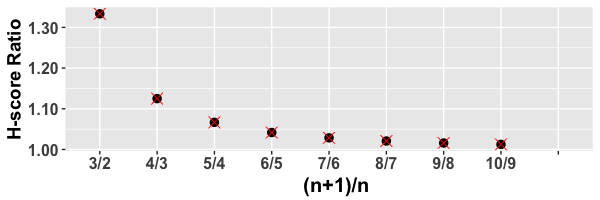}} \hspace{1cm}
	\subfloat[\label{subfig:ratioU1}]{\includegraphics[width=0.46\textwidth]{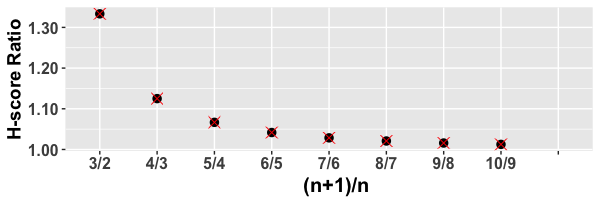}}
	\caption{
		\protect\subref{subfig:HscoreN1}-\protect\subref{subfig:HscoreU1} H-score as function of the number of rows (columns), in a single bicluster of 10 columns (rows). Red dots indicate average H-scores $\overline{H}_{n}$. 
		\protect\subref{subfig:ratioN1}-\protect\subref{subfig:ratioU1} Ratio $r_{n,n+1}$ between average H-scores, with $n=2,\dots,9$ according to simulation (black dot) and Theorem 1 (red cross). 
		The bicluster is generated by an additive model plus noise $a_{ij} = \mu + \varepsilon_{ij}$, in two scenarios with $Var(\varepsilon_{ij})=1$: 
		\protect\subref{subfig:HscoreN1}-\protect\subref{subfig:ratioN1} $\varepsilon_{ij} \sim N(0,1)$; 
		\protect\subref{subfig:HscoreU1}-\protect\subref{subfig:ratioU1} $\varepsilon_{ij}\sim U(-\sqrt{12}/2, \sqrt{12}/2)$.
	}
	\label{fig:example_sd1}
\end{figure}

\begin{figure}[!htb]
	\centering
	\subfloat[\label{subfig:HscoreN2}]{\includegraphics[width=0.45\textwidth]{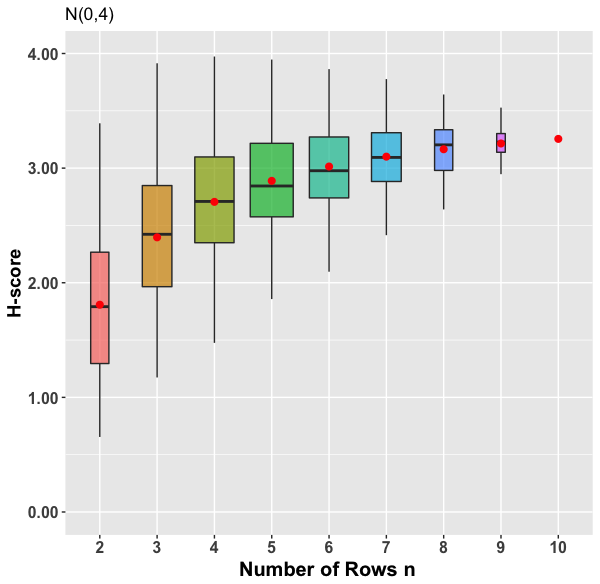}} \hspace{1cm}
	\subfloat[\label{subfig:HscoreU2}]{\includegraphics[width=0.45\textwidth]{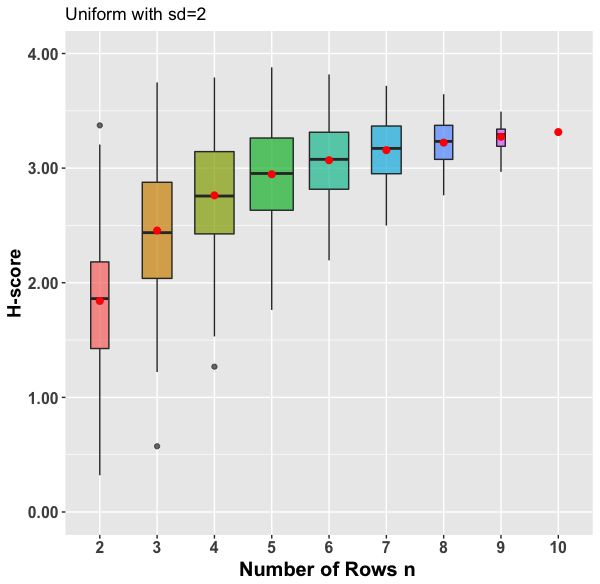}} \\
	\subfloat[\label{subfig:ratioN2}]{\includegraphics[width=0.45\textwidth]{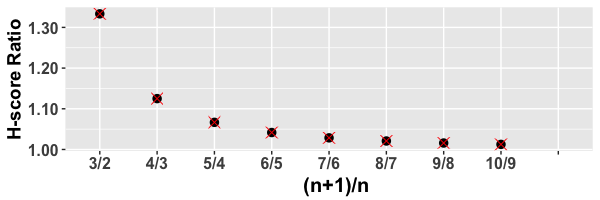}} \hspace{1cm}
	\subfloat[\label{subfig:ratioU2}]{\includegraphics[width=0.45\textwidth]{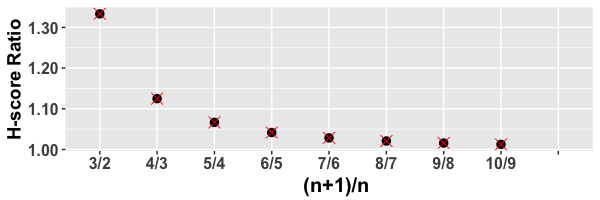}}
	\caption{
		\protect\subref{subfig:HscoreN2}-\protect\subref{subfig:HscoreU2} H-score as function of the number of rows (columns), in a single bicluster of 10 columns (rows). Red dots indicate average H-scores $\overline{H}_{n}$. 
		\protect\subref{subfig:ratioN2}-\protect\subref{subfig:ratioU2} Ratio $r_{n,n+1}$ between average H-scores, with $n=2,\dots,9$ according to simulation (black dot) and Theorem 1 (red cross). 
		The bicluster is generated by an additive model plus noise $a_{ij} = \mu + \varepsilon_{ij}$, in two scenarios with $Var(\varepsilon_{ij})=4$: 
		\protect\subref{subfig:HscoreN2}-\protect\subref{subfig:ratioN2} $\varepsilon_{ij} \sim N(0,4)$; 
		\protect\subref{subfig:HscoreU2}-\protect\subref{subfig:ratioU2} $\varepsilon_{ij}\sim U(-\sqrt{48}/2, \sqrt{48}/2)$.
	}
	\label{fig:example_sd2}
\end{figure}

\noindent The infinite product in (\ref{eq:infprod}) converges, and in particular $r_{2,2+m}\to2$. 
This is due to the fact that $\lim_{i\to\infty} \frac{i^2}{i^2-1} = 1$.
As a consequence, the H-score bias is at most $1$, and it becomes small for comparisons between biclusters of large size (Table \ref{tab:table}). 
We also observe that the value $\overline{H}_{n}$ only depends on the variance of the error term $\varepsilon_{ij}$, and not on the values of $\mu$, $\alpha_i$ and $\beta_j$, nor on the distribution of $\varepsilon_{ij}$ (see Table \ref{tab:table}, Figures \ref{fig:example_sd1}-\ref{fig:example_sd2}). 

\begin{table}[!h]
	\centering
	\begin{tabular}{ccccccccc}
	Model                    & $ \varepsilon_{ij}$     & $\overline{H}_{2}$ & $r_{2,3}$ & $\overline{H}_{3}$ & $r_{3,199}$ & $\overline{H}_{199}$ & $r_{199,200}$ & $\overline{H}_{200}$ \\ \hline
	$\alpha_{i}=0$                    & $\mathcal{N}(0,1)$  & 0.42              & \multirow{2}{*}{1.33}     & 0.57              & \multirow{2}{*}{1.49}       & 0.84                & \multirow{2}{*}{1+3$e^{-5}$}  & 0.84                \\ \cline{2-3}\cline{5-5}\cline{7-7}\cline{9-9}
	$\beta_{j}=0$              & $\mathcal{U}(-2,2)$ & 0.44             &     & 0.58              &        &        0.86              &   &      0.86                \\ \hline
	$\alpha_{i}\neq0 $         & $\mathcal{N}(0,1)$  & 0.42              & \multirow{2}{*}{1.33}     & 0.57              & \multirow{2}{*}{1.49}      & 0.84                &  \multirow{2}{*}{1+3$e^{-5}$}  & 0.84                \\ \cline{2-3}\cline{5-5}\cline{7-7}\cline{9-9}
	$\beta_{j}\neq0$ & $\mathcal{U}(-2,2)$ & 0.44              &      & 0.58              &        &       0.86               &    &          0.86            
	\\ \hline
\end{tabular}
\\ 
\caption{Average H-scores and ratios for different models (with/without row and column differential terms), errors (Gaussian, Uniform) and bicluster sizes.}
\label{tab:table}
\end{table}

\section{Recommendations}
Our results show that employing the H-score to compare biclusters with different numbers of rows or columns could lead to biased results. 
While this bias is small and likely inconsequential for large biclusters, it can be substantial and rather misleading for small biclusters. 
Suppose one is comparing biclusters with $n$ and $n+m$ rows; Equation (\ref{eq:evo}) suggests that this bias can be straightforwardly corrected 
normalizing the H-score ratio by the factor $\prod_{i=n}^{n+m-1}\frac{i^2}{i^2-1}$ 
(an identical reasoning holds for columns).
Notably, this correction should also be employed to adjust the H-score thresholds $\delta$ when finding $\delta$-biclusters. 
Considering the 
seminal role and ubiquitousness of H-scores in biclustering algorithms, and the importance of biclustering algorithms in bioinformatics and computational biology, we believe 
this bias should be taken into serious consideration. The correction we propose is simple to implement, and could help shape the conclusions and insights provided by a broad range of applications.


\section*{Funding}
\small
This work was partially funded by the Eberly College of Science, the Institute for Cyberscience and the Huck Institutes of the Life Sciences (Penn State University);
NSF award DMS-1407639; 
and Tobacco Settlement and CURE funds of the PA Department of Health 
(the Dept.~specifically disclaims responsibility for any analyses, interpretations or conclusions).

\bibliographystyle{Chicago}
\bibliography{references}

\appendix
\section*{Appendix: Proof of Theorem 1}
\begin{proof}
	Given a bicluster $(I,J)$, where $I$ is a set of $N$ rows, and $J$ is a set of $P$ columns, the mean squared residue score is defined as 
	\begin{equation*}
	H(I,J) = \frac{1}{NP} \sum_{i \in I}\sum_{j \in J} (a_{ij} - a_{iJ} - a_{Ij} + a_{IJ})^{2}, 
	\end{equation*}
	
	where we have
	\begin{equation*}
	a_{iJ} = \frac{1}{P} \sum_{j \in J} a_{ij},
	\end{equation*}
	
	\begin{equation*}
	a_{Ij} = \frac{1}{N} \sum_{i \in I} a_{ij},
	\end{equation*}
	
	\begin{equation*}
	a_{IJ} = \frac{1}{NP} \sum_{i \in I}\sum_{j \in J} a_{ij}.
	\end{equation*}
	
	$H(I,J)$ can then be rewritten in the following way:
	\begin{equation*}
	H(I,J) = \frac{1}{NP} \sum_{i \in I}\sum_{j \in J} \left(a_{ij} - \frac{1}{P} \sum_{k \in J} a_{ik} - \frac{1}{N} \sum_{s \in I} a_{sj} + \frac{1}{NP} \sum_{s \in I}\sum_{k \in J} a_{sk}\right)^{2} =
	\frac{1}{NP} \sum_{i \in I}\sum_{j \in J} d_{ij}^2.
	\end{equation*}
	
	Without loss of generality, we have $I=\{1,\dots,N\}$ and $J=\{1,\dots,P\}$, so
	\begin{equation*}
	d_{ij} = a_{ij} - \frac{1}{P}(a_{i1} + \dots + a_{iP}) - \frac{1}{N}(a_{1j} + \dots + a_{Nj}) + \frac{1}{NP}\sum_{s=1}^N\sum_{k=1}^P a_{sk} =
	\end{equation*}
	\begin{align*}
	\frac{NP - N - P + 1}{NP}a_{ij} + \frac{1-N}{NP}\sum_{k \neq j}a_{ik} + \frac{1-P}{NP}\sum_{s \neq i}a_{sj} + \frac{1}{NP}\sum_{s \neq i}\sum_{k \neq j}a_{sk}.
	\end{align*}
	

	Let us notice that $d_{ij}$ is the sum of all the elements in the bicluster, each weighted by a particular coefficient. Then $d_{ij}^2$ is a weighted sum of all the squared elements, and of their double products. Therefore $NP\left[H(I,J)\right]$ is also a weighted sum of all the squared elements in the bicluster, and of their double products.
	
	Let us calculate $h_{i^*j^*}^{(N,P)}$, the coefficient referring to a generic squared element $a_{i^*j^*}^2$ in $NP\left[H(I,J)\right]$. 
	When computing $d_{i^*j^*}^2$, we obtain the coefficient $\frac{(NP - N - P +1)^2}{N^2P^2}$ corresponding to $a_{i^*j^*}^2$. 
	From $d_{i^*j}^2$ with $j \neq j^*$ (same row), we have $\frac{(N-1)^2}{N^2P^2}$, hence $(P-1)\frac{(N-1)^2}{N^2P^2}$ in total. 
	Similarly, from $d_{ij^*}^2$ with $i \neq i^*$ (same column), we have $\frac{(P-1)^2}{N^2P^2}$, hence $(N-1)\frac{(P-1)^2}{N^2P^2}$ in total. 
	Finally, from $d_{ij}^2$ with $i \neq i^*$ and $j \neq j^*$ (elements outside the row $i^*$ and the column $j^*$), we have $\frac{1}{N^2P^2}$, hence $(N-1)(P-1)\frac{1}{N^2P^2}$ in total.
	As a consequence, the coefficient referring to the generic squared element $a_{i^*j^*}$ in $NP\left[H(I,J)\right]$ is
	
	\begin{equation*}
	h_{i^*j^*}^{(N,P)} = \frac{(NP - N - P +1)^2 + (P-1)(N-1)^2 + (P-1)^2(N-1) + (P-1)(N-1)}{N^2P^2}.
	\end{equation*}
	
	Now let us focus on the double products in $NP\left[H(I,J)\right]$. There are three kinds of double products according to the positions of the elements involved in the double product: the case in which two elements belong to the same row, the case in which they belong to the same column, and the case in which they belong to different rows and columns.
	
	\emph{Case 1: same row.} 
	Let $h_{2i^{*} \boldsymbol{\cdot}}^{(N,P)}$ be the coefficient corresponding to the double product of two elements $a_{i^*j_{1}}$ and $a_{i^*j_{2}}$ belonging to the same row $i^*$ and different columns $j_{1} \neq j_{2}$. 
	From $d_{i^*j_1}$ and $d_{i^*j_2}$ we have $4\frac{NP-N-P+1}{NP}\frac{1-N}{NP}$. 
	From $d_{i^*j}$ (same row) with $j \not\in \{j_1,j_2\}$ we get $2\frac{1-N}{NP}\frac{1-N}{NP}$, leading to $2(P-2)\frac{1-N}{NP}\frac{1-N}{NP}$ in total. 
	From $d_{ij_1}$ (same column of the first element) and $d_{ij_2}$ (same column of the second element) with $i \neq i^*$ we get $2\frac{1-P}{NP}\frac{1}{NP}$, leading to $4(N-1)\frac{1-P}{NP}\frac{1}{NP}$ in total. 
	Finally, from all the other $d_{ij}$ with $j \not\in \{j_1,j_2\}$ and $i \neq i^*$ we have $2\frac{1}{NP}\frac{1}{NP}$, for a total of $2(P-2)(N-1)\frac{1}{NP}\frac{1}{NP}$.
	Hence the coefficient is
	
	\begin{equation*}
	h_{2i^{*} \boldsymbol{\cdot}}^{(N,P)} = \frac{4(NP-N-P +1)(1-N) + 2(P-2)(N-1)^2 + 4(1-P)(N-1) + 2(P-2)(N-1)}{N^2P^2}.
	\end{equation*}
	
	\emph{Case 2: same column.} 
	Considering the symmetrical nature of the H-score formulation, the same calculations explained in the case of elements belonging to the same row $i^*$ work for the case of double products of elements belonging to the same column $j^*$. Hence the coefficient $h_{2\boldsymbol{\cdot} j^*}$ referring to the double product of two elements $a_{i_{1}j^*}$ and $a_{i{2}j^*}$ belonging to the same column $j^*$ and different rows $i_{1} \neq i_{2}$ is
	
	\begin{equation*}
	h_{2\boldsymbol{\cdot} j^*}^{(N,P)} = \frac{4(NP-N-P +1)(1-P) + 2(N-2)(P-1)^2 + 4(1-N)(P-1) + 2(N-2)(P-1)}{N^2P^2}.
	\end{equation*}
	
	\emph{Case 3: different rows and columns.}
	Let $h_{2\boldsymbol{\cdot} \boldsymbol{\cdot}}^{(N,P)}$ be the coefficient of the double product of two elements $a_{i_{1}j_{1}}$ and $a_{i_{2}j_{2}}$ which belong to different rows $i_{1} \neq i_{2}$ and columns $j_{1} \neq j_{2}$. 
	From $d_{i_{1}j_{1}}$ and $d_{i_{2}j_{2}}$ we have $4\frac{NP-N-P +1}{NP}\frac{1}{NP}$. 
	From $d_{i_{1}j_{2}}$ and $d_{i_{2}j_{1}}$ we get $4\frac{1-N}{NP}\frac{1-P}{NP}$. 
	From $d_{i_{1}j}$ and $d_{i_{2}j}$ with $j \not\in \{j_{1},j_{2}\}$ (same row as one of the two elements) we have $4\frac{1-N}{NP}\frac{1}{NP}$, leading to $4(P-2)\frac{1-N}{NP}\frac{1}{NP}$ in total; 
	similarly from $d_{ij_{1}}$ and $d_{ij_{2}}$ with $i \neq \{i_{1},i_{2}\}$ (same column as one of the two elements) we have $4\frac{1-P}{NP}\frac{1}{NP}$, leading to $4(N-2)\frac{1-P}{NP}\frac{1}{NP}$ in total. 
	Finally, from all the other $d_{ij}$ with $i \not\in \{i_{1},i_{2}\}$ and $j \not\in \{j_{1},j_{2}\}$ we have $2\frac{1}{NP}\frac{1}{NP}$, for a total of $2(P-2)(N-2)\frac{1}{NP}\frac{1}{NP}$. 
	Hence the coefficient is
	
	\begin{equation*}
	h_{2\boldsymbol{\cdot} \boldsymbol{\cdot}}^{(N,P)} = \frac{4(NP-N-P +1) + 4(1-P)(1-N) + 4(P-2)(1-N) + 4(1-P)(N-2) + 2(P-2)(N-2)}{N^2P^2}.
	\end{equation*}

	Let $H_{n}$ be the H-score of a submatrix of $(I,J)$ composed by $n \leq N$ rows and all the $P$ columns. In a bicluster of $N$ rows there are exactly $\binom{N}{n}$ submatrices with $n$ rows. 
	Let $\overline{H}_{n}$ be their average H-score:
	
	\begin{equation*}
	\overline{H}_{n} = \binom{N}{n}^{-1}\sum_{r=1}^{\binom{N}{n}}H_{n_r},
	\end{equation*}
	
	where $H_{n_r}$ is the H-score of the $r$-th submatrix having $n$ rows and $P$ columns.
	Since each $H_{n_r}$ can be written as a weighted sum of the squared elements belonging to the bicluster and of their double products, so does $\overline{H}_{n}$.
	
	Let us start computing $\overline{h}_{i^*j^*}^{(n)}$, the coefficient referring to a generic squared element $a^2_{i^*j^*}$ in $\overline{H}_{n}$.
	It is useful to notice that the term $a^2_{i^*j^*}$ is present in $H_{n_r}$ if and only if the $r$-th submatrix having $n$ rows and $P$ columns contains the row $i^*$. 
	Of the $\binom{N}{n}$ different submatrices, only $\binom{N-1}{n-1}$ present the row $i^*$.
	Therefore the coefficient $\overline{h}_{i^*j^*}^{(n)}$ is:
	
	\begin{align*}
	\overline{h}_{i^*j^*}^{(n)} 
	&= \frac{\binom{N-1}{n-1}}{\binom{N}{n}} \frac{h_{i^*j^*}^{(n,P)}}{nP} = \\
	&= \frac{\binom{N-1}{n-1}}{\binom{N}{n}} 
	\frac{(nP - n - P + 1)^2 + (P-1)(n-1)^2 + (P-1)^2(n-1) + (P-1)(n-1)}{n^3P^3}  = \\
	&= \frac{(N-1)!}{(n-1)!(N-n)!}\frac{n!(N-n)!}{N!}
	\frac{(P-1)(n-1)nP}{n^3P^3} = \\
	&= \frac{n}{N} \frac{(P-1)(n-1)}{n^2P^2} = \\
	&= \frac{(P-1)(n-1)}{NnP^{2}}.
	\end{align*}
	
	Now let $\overline{h}_{2i^{*} \boldsymbol{\cdot}}^{(n)}$ be the coefficient in $\overline{H}_n$ referring to the double product of two elements $a_{i^*j_1}$ and $a_{i^*j_2}$ belonging to the same row $i^*$ and different columns $j_1 \neq j_2$. Considering the fact that $a_{i^*j_1}$ and $a_{i^*j_2}$ belong to the same row $i^*$, there are exactly $\binom{N-1}{n-1}$ submatrices composed by $n$ rows having the row $i^*$. 
	Therefore the coefficient is:
	
	\begin{align*}
	\overline{h}_{2i^{*} \boldsymbol{\cdot}}^{(n)} 
	&= \frac{\binom{N-1}{n-1}}{\binom{N}{n}} 
	\frac{h_{2i^{*} \boldsymbol{\cdot}}^{(n,P)}}{nP} = \\
	&= \frac{\binom{N-1}{n-1}}{\binom{N}{n}} 
	\frac{4(nP-n-P +1)(1-n) + 2(P-2)(n-1)^2 + 4(1-P)(n-1) + 2(P-2)(n-1)}{n^3P^3} = \\
	&= \frac{n}{N} \frac{2(1-n)}{n^2P^2} = \\ 
	&= \frac{2(1-n)}{NnP^2}
	\end{align*}
	
	Now let $\overline{h}_{2\boldsymbol{\cdot}j^*}^{(n)}$ be the coefficient in $\overline{H}_n$ referring to the double product of two elements $a_{i_1j^*}$ and $a_{i_2j^*}$ belonging to the same column $j^*$ and different rows $i_1 \neq i_2$. Since $a_{i_1j^*}$ and $a_{i_2j^*}$ do not belong to the same row, there are exactly $\binom{N-2}{n-2}$ submatrices composed by $n$ rows having both row $i_1$ and row $i_2$. 
	Hence the coefficient is:
	\begin{align*}
	\overline{h}_{2\boldsymbol{\cdot}j^*}^{(n)} 
	&= \frac{\binom{N-2}{n-2}}{\binom{N}{n}}
	\frac{h_{2\boldsymbol{\cdot}j^*}^{(n,P)}}{nP} = \\
	&= \frac{\binom{N-2}{n-2}}{\binom{N}{n}}
	\frac{4(nP-n-P +1)(1-P) + 2(n-2)(P-1)^2 + 4(1-n)(P-1) + 2(n-2)(P-1)}{n^3P^3} = \\
	&= \frac{n(n-1)}{N(N-1)} \frac{2(1-P)}{n^2P^2} = \\
	&= \frac{2(1-P)(n-1)}{NnP^2(N-1)}
	\end{align*}
	
	Finally, let $\overline{h}_{2\boldsymbol{\cdot} \boldsymbol{\cdot}}^{(n)}$ be the coefficient in $\overline{H}_n$ referring to the double product of two elements $a_{i_1j_1}$ and $a_{i_2j_2}$ which belong to different rows $i_1 \neq i_2$ and columns $j_1 \neq j_2$. 
	Since $a_{i_1j_1}$ and $a_{i_2j_2}$ do not belong to the same row, there are exactly $\binom{N-2}{n-2}$ submatrices composed by $n$ rows presenting both row $i_1$ and row $i_2$. 
	Therefore the coefficient is:
	
	\begin{align*}
	\overline{h}_{2\boldsymbol{\cdot} \boldsymbol{\cdot}}^{(n)}  
	&= \frac{\binom{N-2}{n-2}}{\binom{N}{n}}
	\frac{h_{2\boldsymbol{\cdot} \boldsymbol{\cdot}}^{(n,P)}}{nP} = \\
	&= \frac{\binom{N-2}{n-2}}{\binom{N}{n}}
	\frac{4(nP-n-P +1) + 4(1-P)(1-n) + 4(P-2)(1-n) + 4(1-P)(n-2) + 2(P-2)(n-2)}{n^3P^3} = \\
	&= \frac{n(n-1)}{N(N-1)} \frac{2}{n^2P^2} = \\ 
	&= \frac{2(n-1)}{NnP^2(N-1)}
	\end{align*}

	Being interested on the relationship between $\overline{H}_{n+1}$ and $\overline{H}_n$, we compute
	$\overline{h}_{i^*j^*}^{(n+1)}$ (i.e. the coefficient of the generic squared element $a_{i^*j^*}^2$ in $\overline{H}_{n+1}$),
	$\overline{h}_{2i^{*} \boldsymbol{\cdot}}^{(n+1)}$ (i.e. the coefficient referring to the double product of two elements belonging to the same row $i^*$ in $\overline{H}_{n+1}$), 
	$overline{h}_{2\boldsymbol{\cdot}j^*}^{(n+1)}$ (i.e. the coefficient referring to the double product of two elements belonging to the same column $j^*$ in $\overline{H}_{n+1}$), and
	$\overline{h}_{2\boldsymbol{\cdot} \boldsymbol{\cdot}}^{(n+1)}$ (i.e. the coefficient referring to the double product of two elements which belong to different rows and columns in $\overline{H}_{n+1}$). We have:
	
	\begin{align*}
	\overline{h}_{i^*j^*}^{(n+1)} &= \frac{(P-1)n}{N(n+1)P^{2}}; \\
	\overline{h}_{2i^{*} \boldsymbol{\cdot}}^{(n+1)} &= \frac{-2n}{N(n+1)P^2}; \\
	\overline{h}_{2\boldsymbol{\cdot}j^*}^{(n+1)} &= \frac{2(1-P)n}{N(n+1)P^2(N-1)}; \\
	\overline{h}_{2\boldsymbol{\cdot} \boldsymbol{\cdot}}^{(n+1)} &= \frac{2n}{N(n+1)P^2(N-1)}.
	\end{align*}
	
	Then we have:
	
	\begin{align*}
	\frac{\overline{h}_{_{i^*j^*}}^{(n+1)}}{\overline{h}_{_{i^*j^*}}^{(n)}} 
	&= 
	\frac{(P-1)n}{N(n+1)P^{2}}\frac{NnP^{2}}{(P-1)(n-1)}= \frac{ n^{2}}{n^{2} -1}; \\
	\frac{\overline{h}_{2i^{*} \boldsymbol{\cdot}}^{(n+1)}}{\overline{h}_{2i^{*} \boldsymbol{\cdot}}^{(n)} }
	&= 
	\frac{-2n}{(n+1)P^2N} \frac{nP^2N}{2(1-n)} = \frac{n^2}{n^2-1}; \\
	\frac{\overline{h}_{2 \boldsymbol{\cdot}j^{*}}^{(n+1)}}{\overline{h}_{2 \boldsymbol{\cdot}j^{*}}^{(n)} }
	&= 
	\frac{2(1-P)n}{N(n+1)P^2(N-1)} \frac{NnP^2(N-1)}{2(1-P)(n-1)} = \frac{n^2}{n^2-1}; \\
	\frac{\overline{h}_{a_{i^*j^*}a_{\hat{i}\hat{j}}}^{n+1}}{\overline{h}_{a_{i^*j^*}a_{\hat{i}\hat{j}}}^{n}} 
	&= 
	\frac{2n}{(n+1)P^2N(N-1)}\frac{nP^{2}N(N-1)}{2(n-1)} = \frac{n^2}{n^2-1}.
	\end{align*}
	
	Now let us consider $\overline{H}_{n}$ and $\overline{H}_{n+1}$ with $n \leq N$. Both are weighted sums of squared elements of the bicluster and their double products. Being the ratio of each pair of coefficients in $\overline{H}_{n+1}$ and $\overline{H}_{n}$ equal to $\frac{n^2}{n^2-1}$, then:
	
	\begin{equation*}
	\overline{H}_{n+1} = \frac{n^2}{n^2-1} \overline{H}_{n}.
	\end{equation*}

	Analogously, considering the average H-score $\overline{H_{p}}$ of all the submatrices of $(I,J)$ composed by $N$ rows and $p \leq P$ columns, we obtain:
	
	\begin{equation*}
	\overline{H}_{p+1} = \frac{p^2}{p^2-1} \overline{H}_{p}.
	\end{equation*}
\end{proof}

\end{document}